\documentclass[aps,pra,twocolumn,floatfix,showpacs]{revtex4}

\usepackage{amssymb}
\usepackage{amsmath}
\usepackage{graphicx}
\usepackage{amssymb}

\newcommand{\unit}[1]{\rm{#1}}
\newcounter{oldenumi}

\begin{document}

\preprint{}
\title[ShortTitle]{ Charge qubits and limitations of electrostatic quantum gates}
\author{A.~Weichselbaum}
\author{S.~E.~Ulloa}
\affiliation{Department of Physics and Astronomy, Nanoscale and
Quantum Phenomena Institute, Ohio University, Athens, Ohio
45701-2979} 

\begin{abstract}
We investigate the characteristics of purely electrostatic
interactions with external gates in constructing full single qubit
manipulations. The quantum bit is naturally encoded in the spatial
wave function of the electron system. Single-electron--transistor
arrays based on quantum dots or insulating interfaces typically
allow for electrostatic controls where the inter-island tunneling
is considered constant, e.g. determined by the thickness of an
insulating layer. A representative array of $3\times 3$ quantum
dots with two mobile electrons is analyzed using a Hubbard
Hamiltonian and a capacitance matrix formalism. Our study shows
that it is easy to realize the first quantum gate for single qubit
operations, but that a second quantum gate only comes at the cost
of compromising the low-energy two-level system needed to encode
the qubit. We use perturbative arguments and the Feshbach
formalism to show that the compromising of the two-level system is
a rather general feature for electrostatically interacting qubits
and is not just related to the specific details of the system
chosen.  We show further that full implementation requires tunable
tunneling or external magnetic fields.
\end{abstract}

\volumeyear{year}
\volumenumber{number}
\issuenumber{number}
\eid{identifier}
\date[Date: ]{22 Dec 2003}
\maketitle

\section{Introduction}

Quantum computation in its binary concept requires a set of two different
quantum states, a \textit{quantum two-level system} (qu2LS), that realizes
the quantum bit physically \cite{Nielsen00}. Some physical systems are
intrinsically qu2LSs such as the spin $1/2$ of a fermion or the polarization
of a photon. Spin $1/2$ systems such as some nuclei of atoms or electrons
are considered well-suited also because of the comparatively long spin
decoherence times ($\sim \unit{ms}$ to $\unit{ns}$) \cite%
{Levy01,Gordon58,Kikkawa97} which derives from the rather weak
interaction of the spin with its environment. This long coherence
also has the price that important processes in qubit operation are
rather slow processes. On the other hand, decoherence times for
charge based quantum systems are orders of magnitude shorter
($\sim\unit{ns}$) due to the comparatively strong Coulomb
interaction \cite{Hayashi03,Storcz03}. Yet, all quantum gates
including interacting qubits can be expected to scale similarly in
time and to allow for fast qubit operations well below the
decoherence limits. Furthermore, final readout through
single-electron transistors (SET) or quantum point contacts (QPC)
appears to be ``straightforward" (i.e., implementable in
principle) \cite{Goan01,DiCarlo03,Devoret00}.

In this paper we discuss quantum bits (qubits) encoded in the
spatial wave function of electrons embedded in condensed matter
systems, and the emphasis is placed on whether it is possible to
realize a qubit based solely on charge distribution
(\textit{charge qubit}) and capacitive coupling. For example,
having a set of two quantum dots (qudots) close enough so that one
electron can tunnel back and forth, one may envisage a qubit where
the electron being on one quantum dot (qudot) represents one
state, and being on another qudot the other state
\cite{Hayashi03}. Other examples are the cellular automata setups
with a $2\times 2$ array of qudots with excess electrons
\cite{Toth01,Gardelis03}. These proposals are fundamentally based
on the variability of the tunneling $t$. Yet there are many
physical quantum systems where the handle on the tunneling is
limited or non-existent, such as the case of metallic qudot
structures where the tunneling barrier is determined by the
thickness of oxide layers in the structure.

This then suggests the question of whether there is a way that
electrostatically controlled logical quantum gates (qugates) can
realize the necessary single qubit operations and the tunability
of interactions between them. In this context, it is important to
clearly define what one means by a quantum bit encoded in a
quantum two-level system (qu2LS) and what are the requirements for
it. The following criteria are established for the usefulness of a
qu2LS \cite{Nielsen00}:

\renewcommand{\theenumi}{A\arabic{enumi}} \renewcommand{\labelenumi}{%
\theenumi.}

\begin{enumerate}
\item \label{qubit_criteria_1} The qu2LS should include the ground
state of the system in the \textit{working-range} of the (tunable)
parameters; this clearly facilitates the initialization process in
an experiment and is much more reliable when compared to a qu2LS
completely built on excited states.

\item \label{qubit_criteria_2} The qu2LS should be well separated
in energy from the remaining states in the Hilbert space. This
reduces the influence of the remaining Hilbert space, whose
interference can be insofar interpreted as a source of
\textit{decoherence} and transitions to which may result in loss
of probability in the primary system. This \textit{lossy channel}
for the qu2LS is considered in more detail below.

\item \label{qubit_criteria_3} The qu2LS must interact with a set
of external gates in order to control single qubit states as well
as the interaction between them without compromising the two-level
system.
\end{enumerate}

In the case of charge located on a set of well-defined quantum
dots (a \textit{qudot network}), the electrostatic interaction is
clearly able to satisfy point (\ref{qubit_criteria_3}) where
Coulomb blockade (or charging) effects introduce a high energy
scale ($\gtrsim 1\unit{meV}$) in typical qudots. These structures
localize the operating electrons and limit the unwanted fast
decoherence times due to interaction with the surrounding
condensed matter environment. Typical decoherence sources such as
phonons or unstable impurities in the environment are regarded
frozen out or static, respectively, at the low temperatures
required for operation of the qubit system. Consequently, the
primary source of decoherence in this regime is in fact the
reservoir of high-lying states, and its coherent interference with
the qu2LS will be consider explicitly in our description.

The effect of the controlling electrostatic gates on the
\textit{reservoir }of high-lying states must be in the adiabatic
regime. The time dependent manipulation through gate action can be
estimated in the following manner: the gates are considered to act
in clearly specified time windows in a step like behavior: they
are turned on and off at will.  This switching, however, is always
carried out with a maximum speed which introduces a characteristic
frequency $\omega _{switch}=2\pi /\tau _{switch}$, where $\tau
_{switch}$ is the switching time itself. In order for the
influence of the reservoir of higher lying states to be
negligible, $\hbar \omega _{switch}$ must be much smaller than the
energy difference to the closest coupled states in that bath; thus
the switching must be done smoothly enough (adiabatically), so as
to not admix higher states into the lower qubit states. Coherent
quantum operation in the qu2LS, however, demands the switching to
be done faster than $1/\delta $, where $\delta $ is the splitting
of the qu2LS in question, and incorporated in criterion
(\ref{qubit_criteria_2}).

With respect to criteria (\ref{qubit_criteria_2}) and
(\ref{qubit_criteria_3}), an estimate of how much of the wave
function may be lost for each gate operation can be obtained from
the Feshbach formalism \cite{Feshbach62}. For the unperturbed
low-energy state manifold $\mathcal{P}$ (in contrast to the
remainder of the space $\mathcal{Q}$), an initial state $|\psi
\rangle $ fully contained in $\mathcal{P}$ will acquire
projections in $\mathcal{Q}$ due to a gate operation $V$ (assumed
instantaneous), given by
\begin{equation*}
\langle \psi _{Q}|\psi _{Q}\rangle =\left\langle \psi
_{P}\right\vert H_{PQ} \frac{1}{E-H_{QQ}}\cdot
\frac{1}{E-H_{QQ}}H_{QP}\left\vert \psi _{P}\right\rangle \, ,
\end{equation*}%
where $H=H_{0}+V$ and $H_{PQ}\equiv PHQ$ is a projection of the
Hamiltonian, and with the other projections defined similarly (see
Eq.~(\ref{H_PQ_rep}) below). Space $\mathcal{Q}$ is considered to
be at least an energy $\Delta _{0}$ separated from space
$\mathcal{P}$ and the change in the matrix elements due to the
gate operation $V$ is approximated by the splitting $\delta $
induced by that very $V$ in the ground state pair in
$\mathcal{P}$; with this, the equation above can be estimated as
\begin{equation}
\langle \psi _{Q}|\psi _{Q}\rangle \lesssim \left\langle \psi
_{P}\right\vert \left( \frac{\delta }{\Delta _{0}}\right)
^{2}\left\vert \psi _{P}\right\rangle =\left( \frac{\delta
}{\Delta _{0}}\right) ^{2} \, . \label{loss_through_switching}
\end{equation}%
The gate operations considered are a sequence of steps in the
external parameters, which implies that with every one of these
steps a small probability fraction is lost from the ground qu2LS
to the remaining higher lying states, and as such it can be
considered as an additional channel for decoherence, even if the
projection is nearly reversible in a gate cycle. Moreover, if the
ground state can be considered sufficiently isolated ($ \delta \ll%
\Delta $), the error drops quadratically with the ratio $ \delta%
/\Delta $, so that if $\delta \simeq 0.03\Delta $, the probability
lost per gate operation would be smaller than $0.1~\%$. However,
because the gate operations will never be performed
instantaneously -- smoothing the transitions so that they take
longer than $\Delta _{0}^{-1}$ but are faster than $\delta ^{-1}$
-- this is clearly to reduce the probability loss to the
`$\mathcal{Q}$ reservoir' of excited states.

Thus with proper adiabatic design of the qugates with respect to
the higher lying `reservoir', the way to single qubit operations
is open. However, how exactly these are realized still leaves
plenty of possibilities, which one can imagine being flexible
enough. Here the main emphasis is placed on capacitively coupled
quantum gates and the question of whether they allow the necessary
single qubit operations. Most surprisingly, the answer will turn
out negative. Despite the great degree of flexibility in geometry
of electrostatic gates and system design, we will show below that
it is not possible to implement fully operational qugates without
compromising the robustness of the qubits.

\section{The Model System}

The model network under consideration is a $3\times 3$ array of
qudots with a single state per site and spin included. For
theoretical purposes the structure is taken large enough to
illustrate the main physics. For an experiment in this area,
however, it is likely more practical to choose an array with fewer
dots and gates. For the analysis, the $3\times 3$ array is
flexible and manageable, and is used to illustrate our generalized
conclusions. Other geometries have been explored and yield similar
results.

\begin{figure}[tbp]
\includegraphics*[angle=0,width=8.6cm]{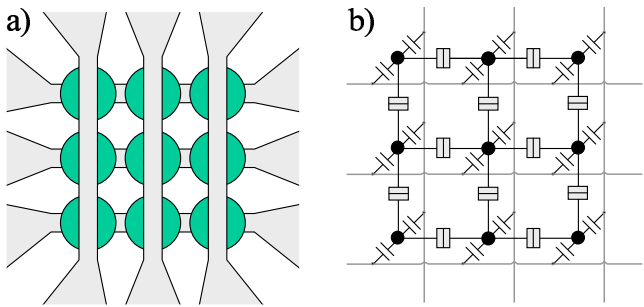}
\caption{Model system - $3\times 3$ array: (a) schematic layout:
circles represent quantum dots and the three horizontal and three
vertical bars represent the gates which are connected to the
outside world. (b), same as panel (a) but drawn as a capacitor and
tunnel junction network, where black circles represent the qudots.
The box symbols in between dots represent capacitive tunnel
junctions.} \label{fig_3x3_setup}
\end{figure}

The $3\times 3$ array of qudots is sketched together with the set
of external gates in Fig.~\ref{fig_3x3_setup}a. As seen from panel
(b), only nearest neighbor capacitances are taken into account as
well as nearest neighbor tunneling between dots. Overall, four
parameters enter the model: the capacitance from each dot to
either one of the gates ($C_{g}=45 \unit{aF}$), the
nearest-neighbor dot-dot capacitance ($C_{dd}=45\unit{aF}$), the
dot self capacitance ($C_{d0}=45\unit{aF}$) and the nearest
neighbor dot-dot tunneling ($t\sim 2\unit{\mu eV}$). For these
values, the energy cost for double occupancy (the Hubbard $U$)
becomes $U=1~\unit{meV}$, a typically used value in this context.
Including a dielectric constant of $\epsilon =10$, these numbers
represent a typical design where dots of size $200\unit{nm}$ are
separated by approximately also $200\unit{nm}$. These numbers can
of course change depending on the detailed geometry and shapes of
the dots. In that sense, the circles in Fig.~\ref{fig_3x3_setup}a
are just a symbolic representation of the dots. However, the
capacitance values chosen correspond to relatively large
nano-structures, and the resulting low energy and temperature
requirements of few tens of $\unit{mK}$ may thus be lifted to some
extent by going to smaller structures.

The Hamiltonian used to describe this system is of the extended
Hubbard type
\begin{eqnarray}
H &=&\sum_{i,\sigma }\varepsilon _{\sigma }\,c_{i\sigma
}^{+}c_{i\sigma }-\sum_{i,j,\sigma }t_{ij}^{\sigma }(c_{i\sigma
}^{+}c_{j\sigma }+c_{j\sigma }^{+}c_{i\sigma })+
\notag \\
&&\frac{1}{2}\sum_{i,j}V_{ij\,}\hat{n}_{i}\hat{n}_{j}+\sum_{i}V_{i\,}\hat{n}%
_{i}\text{,} \label{Hubbard_H}
\end{eqnarray}%
where $\hat{n}_{i}\equiv c_{i\uparrow }^{+}c_{i\uparrow }+c_{i\downarrow
}^{+}c_{i\downarrow }$ in the standard notation. The $\varepsilon
_{(i)\sigma }$ refers to the local energy of the state $\sigma $ on the $%
i=\{1,\ldots ,n\}$ identical dots and can be used to account for the Zeeman
splitting of spins in an external magnetic field. For most cases, $%
\varepsilon _{\sigma }$ is simply set to zero. The tunneling coefficients $%
t_{ij}^{\sigma }$ are considered independent of the spin orientation, thus $%
t_{ij}^{\uparrow }=t_{ij}^{\downarrow }\equiv t_{ij}$, and unequal
to zero only between nearest neighbors in the qudot network. The
electrostatic energy in the last two terms of
Eq.~(\ref{Hubbard_H}), i.e. the coefficients $V_{ij}$ and $V_{i}$,
are derived from the (total) capacitance matrix of the system (see
App.~\ref{appendix_C-Matrix} for more details).

Throughout this paper, the electronic system of qudots is
considered to have, for simplicity, a fixed number of two
electrons ($2e$) operating in it. Furthermore, since spin flip
processes occur on a comparatively long time scale
\cite{Levy01,Gordon58,Kikkawa97}, they are neglected and thus the
overall spin is considered constant. The correct statistics for
exchange of the two electrons is taken into account by the
fermionic creation (annihilation) operators $c_{i\sigma }^{+}$
($c_{i\sigma } $). Since spin is conserved, the basis for the two
spin $1/2$ particles ($2e$) is conveniently changed to singlet and
triplet states which are also spin eigenstates. Fermionic
statistics constrains the spatial wavefunction of the (triplet)
singlet states under particle exchange to be (anti)symmetric,
respectively. Henceforth, the spin index can be dropped
completely, and the distinction between singlet and triplet states
can be incorporated by using (fermionic) bosonic operators for
(triplet) singlet states, respectively, with the constraint of two
particle occupation in the system.

\section{Single Qubit Quantum Gates}

\begin{figure}[tbp]
\includegraphics*[angle=0,width=8.6cm]{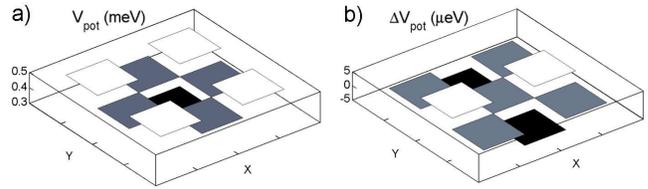}
\caption{Potential landscape for array in
Fig.~\protect\ref{fig_3x3_setup}.  (a) Single particle potential
on array with no gate voltages applied ($\mathbf{V}_{g}=0$). (b)
Change of single particle potential due to one specific set of
applied gate voltages.} \label{fig_3x3_vpot}
\end{figure}

The $3\times 3$ model introduced in the previous section is
representative for a network with $C_{4v}$ symmetry. With this
geometrical symmetry, several states will have a natural $2$-fold
degeneracy. Having set up the total capacitance matrix for the
system, the single particle potential landscape for the array is
shown in Fig.~\ref{fig_3x3_vpot}a for the case of no gate voltages
applied. The absence of nearest neighbors on the outer boundary
results in an electrostatic potential well such that a single
charge stays preferentially in the center of the array. Applying a
peculiar pattern of gate voltages, the $90^{\circ }$ symmetry of
the potential can be broken which effectively alters the potential
on the middle outer islands only; Fig.\ \ref{fig_3x3_vpot}b
depicts an example of such a situation.

\subsection{Numerical Simulations}

\begin{figure}[tbp]
\includegraphics*[angle=0,width=8.6cm]{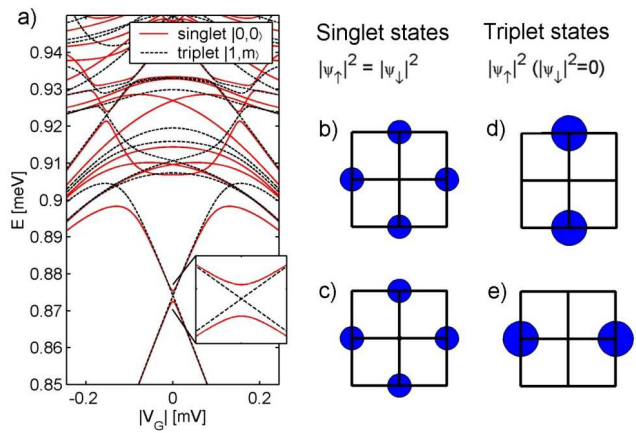}
\caption{(a) Energy level spectrum of the $3\times 3$ system and
dependence of the symmetry breaking pattern of gate voltages in
Fig.\ \ref{fig_3x3_vpot}b.
Singlet states $\left\vert s,s_{z}\right\rangle =\left\vert%
0,0\right\rangle $ are shown in red, while triplet states $%
\left\vert s,s_{z}\right\rangle =\left\vert%
1,m=\{+1,0,-1\}\right\rangle $ are shown in dashed black. (b) and
(c) Probability distribution over the $3\times 3$ array of the
ground pair (qu2LS) for singlet states.  Notice equal probability
for spin up and spin down ($|\protect\psi _{\uparrow%
}|^{2}=|\protect\psi _{\downarrow }|^{2}$).  (d) and (e) Lowest
triplet states.  Case chosen ($s_{z}=+1$) has only a spin up
component (note, however, that spatial probability distribution is
the same for the sixfold degenerate triplet states).}
\label{fig_3x3_level_spectrum}
\end{figure}

Adding a second charge to the $3\times 3$ array leads to a
competing effect of the potential well structure in
Fig.~\ref{fig_3x3_vpot}a with Coulomb repulsion and the two
charges push each other halfway outside (see Fig.~\ref
{fig_3x3_level_spectrum}d+e) resulting in two low-lying
classically degenerate states with charges arranged either
vertically (\ref {fig_3x3_level_spectrum}d) or horizontally
(\ref{fig_3x3_level_spectrum}e). The eigenspectrum for the
$3\times 3$ system with two electrons and its dependence on the
gate voltage pattern shown in Fig.~\ref{fig_3x3_vpot}b is plotted
in Fig.~\ref{fig_3x3_level_spectrum}a. The three different spin
configurations for triplet states have exactly the same
eigenspectrum. So looking at one specific triplet spin
configuration, there is still an exact degeneracy in the ground
state due to the spatial symmetry when no gate voltages are
applied. The corresponding spatial configurations are shown in
Fig.~\ref{fig_3x3_level_spectrum}d+e. For the singlet states,
however, a gap opens up. This gap originates from the distinct
exchange symmetry in the spatial part of the wave function when
compared to the triplet states. Taking as basis states those shown
for the triplet manifold in Fig.~\ref{fig_3x3_level_spectrum}d+e,
it is clear that for the singlet subspace they mix into their
symmetric and antisymmetric combinations (bonding/antibonding
states) near the degeneracy point where no gate voltages are
applied ($\mathbf{V}_{g}=0$). One can consider the tunneling part
of the Hamiltonian as a perturbation ($H=H_{0}+V\left( t\right)
$), so that the extra minus sign for particle exchange in the
spatial part of the triplet wave functions has an interesting
effect: the perturbative terms that mix the two basis states
effectively cancel out, so that the degeneracy stays intact. The
underlying reason, as seen below, is that due to the $C_{4v}$
symmetry for each path there also exists a mirrored path where the
particles are exchanged. However, for the singlet states, the
perturbative terms all come with the same sign and add up. Thus
the basis states effectively mix, resulting in the singlet states
shown in Fig.~\ref{fig_3x3_level_spectrum}b+c. The gap that opens
then for the singlet states as a function of gate voltages, is
properly described as an anticrossing in the level spectrum and
this immediately opens the path towards generating Rabi
oscillations in the system characterized by different charge
configurations.

The gap between singlet states with no voltages applied ($\mathbf{%
V}_g=0$) can be estimated using the Feshbach formalism
\cite{Feshbach62} which is a well-suited approximation for a state
space well separated from the remainder of the Hilbert space and
can be thought of as a perturbative approach that builds on path
histories in the Hilbert space. The lowest order term for weak
tunneling $t$ and consistent with the numerical data is given as
\begin{equation}
\delta \sim 32\frac{t^{4}}{\Delta _{0}^{3}}\ \text{,}
\label{delta_level_splitting}
\end{equation}%
where $\Delta_0$ is the energy gap to the excited manifold
($\Delta_0\simeq 0.03\unit{meV}$ in Fig.\ 3a). This result can be
visualized as four hops ($t^{4}$) needed at the cost of at least
$\Delta _{0}$ for each of the three intermediate states ($\Delta
_{0}^{-3}$); further, the prefactor gives the number of possible
low energy paths to go from one basis state to the other including
particle exchange symmetry.

\begin{figure}[tbp]
\includegraphics*[angle=0,width=8.6cm]{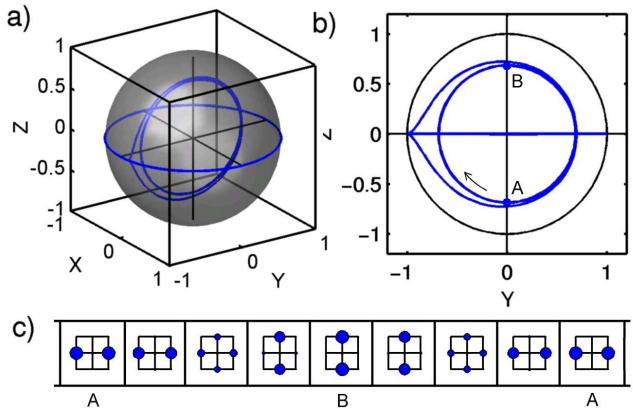}
\caption{Coherent manipulation of the singlet state under gate
action - (a) Evolution of the qubit in the Bloch sphere
representation after projection onto the basis of the (initial)
eigenspectrum at $\mathbf{V}_{g}=0$. The Bloch sphere is shown in
black, and the evolution of the Bloch vector in the qu2LS is shown
in blue. (b) Same as panel (a), but side view, showing the slight
size reduction due to the adiabatic interaction with the higher
lying reservoir of states. (c) Coherent Rabi oscillations for the
sequence $A\rightarrow B\rightarrow A$ in panel (b) in the
direction indicated with $t=10\unit{\mu eV}$. The probability
distribution in real space is shown for the $3\times 3 $ array
over equally spaced time intervals in a total time window of $0.56
\unit{ns}$ which corresponds to one period for this tunneling
based action.} \label{fig_3x3_bloch_sphere}
\end{figure}

A numerical simulation of the state evolution that makes use of
the anticrossing of the singlet state is shown in
Fig.~\ref{fig_3x3_bloch_sphere}. The state of the qu2LS is
represented by a three dimensional vector in the Bloch sphere
\cite{Nielsen00} where the (initial) eigenstates for $\mathbf{V}%
_{g}=0$ are taken as the basis for this representation. Note that
this is a slightly modified definition of the Bloch vector,
insofar as there is a reservoir of higher lying states accessible
to the system. Panels (a+b) show the \textit{path }of the Bloch
vector as three sequential qubit operations were performed: the
first and the last operation are based on the symmetry breaking
gate voltage pattern shown in Fig.~\ref{fig_3x3_vpot}b, which
rotate the Bloch vector around the $z$ axis, the horizontal
circular path in panel (a). For demonstrational purposes, the
second operation is based on tunable tunneling (an effective
$\sigma _{x}$ gate in the pseudo-spin space of the qubit, see
below). This second operation rotates the Bloch vector around the
$x$-axis and together with the first operation allows one to
rotate the Bloch vector anywhere in the Bloch sphere as required
for single qubit operations. Panel (c) shows the evolution of the
real space probability distribution during the
second operation. The two basis states in Fig.~%
\ref{fig_3x3_level_spectrum}d+e are nicely rotated into each other
over a time consistent with the nature of the Rabi oscillations.
Typical rise-times for the voltage gate that do not mix in higher
lying states are well below the 1~ps range, while if one were to
tune the tunneling, the minimal rise-times for adiabatic switching
would require times in the $100\unit{ps}$ range, in order to limit
the admixing of higher lying states. The adiabatic regime
considered here is clearly seen in the Bloch sphere representation
of Fig.~\ref{fig_3x3_bloch_sphere}b by observing that the length
of the Bloch vector is reduced to less than one in the
intermediate gate operation.  On return to the initial parameters,
this amplitude temporarily lost to the bath is regained, however.

\subsection{Rabi Oscillations and 2nd Quantum Gate}

Qubits are conveniently mapped onto the spin $1/2$ formalism using
Pauli matrices \cite{Nielsen00, Ballentine99}; the system is
described by a pseudospin which can be rotated in $3D$ space by
applying perturbations which effectively act as magnetic field
along different directions (note that there is no real magnetic
field and that the real spin of the two electron system is taken
care of by the singlet and triplet states). An arbitrary single
qubit operation thus requires the realization of two distinct
rotations in the $3D$ pseudo-spinor space. This translates to two
linearly independent combinations of the three Pauli matrices
$\sigma _{\left\{ x,y,z\right\} }$ required to implement the
necessary quantum gates. $\sigma _{z}$ is easily implemented using
the gate voltages and thus applying different potentials to
different regions in the qudot array as outlined in the previous
section. Further, if one considers the tunneling $t$ to be
approximately constant, like in lithographically grown qudot
structures with the tunneling determined by oxide layers, one may
ask if it is possible to obtain the second qugate by applying a
peculiar pattern of only capacitively coupled voltage gates. The
answer turns out negative in the sense that either the second gate
($\sigma _{x}$) is orders of magnitude weaker than the first gate
($\sigma _{z}$), or it compromises the two-level ground state
system such that at least one initially well split off eigenstate
comes within gap distance to the qubit encoding subspace
\cite{note}.

\begin{figure}[tbp]
\includegraphics*[angle=0,width=8.6cm]{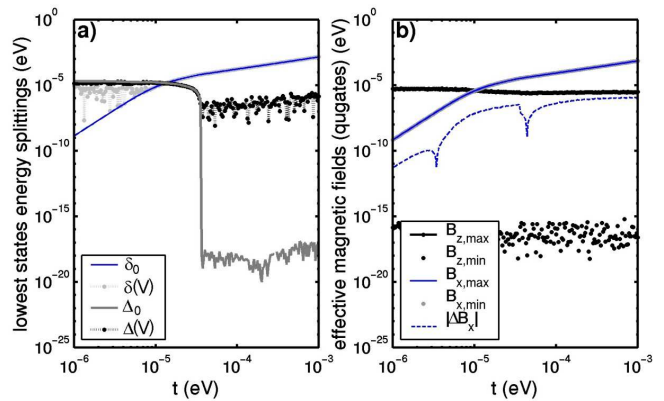}
\caption{Numerical exploration of effective (pseudo-) magnetic
fields from a random sequence of gate voltages (4096
configurations for every $t$ value). (a) Energy level splitting
between the lowest two eigenstates (the qu2LS), $\delta$, as well
as the level splitting between the 2nd and the 3rd eigenstate,
$\Delta$, shown with and without gate voltages applied. A well
behaved two
level system exists for $t \protect\lesssim 0.5 \times%
10^{-5}\unit{eV}$, while for larger $t$ higher lying states cross
over. (b) Sampling the gate voltages randomly, the minimum and
maximum pseudo-magnetic fields achieved are recorded
($H=a1+\mathcal{\vec{B}}\vec{\protect \sigma}$, and thus
$\mathcal{\vec{B}}$ has units of energy). Since $\mathcal{%
B}_{x,min}$ and $\mathcal{B}_{x,max}$ are very similar, the
difference $\Delta \mathcal{B}_{x}$ is shown explicitly by the
blue dashed line.  $\mathcal{B}_{z,min/max}$ values are clearly
discernible. Note that the $\mathcal{B}_{x}$ is directly related
to the gap in the ground state ($\delta _0$ in panel (a))}
\label{fig_3x3_tbeff}
\end{figure}

As a way to illustrate this result in the $3\times 3$ system of
Fig.~\ref{fig_3x3_setup}a, all six gate voltages were sampled
randomly within their parameter space and over a significant range
of tunneling coefficient $t$ values. The results are shown in
Fig.~\ref{fig_3x3_tbeff}. From panel (a), the region of the intact
two-level system involving the ground state is identified as the
region where $\delta \left( V\right) \ll \Delta \left( V\right) $
and thus $t\leq 5\unit{\mu eV}$; $\delta \equiv E_1 - E_0$ and
$\Delta \equiv E_2 - E_1$ are the energy differences of the lowest
three states. It is then clear that the condition $\delta\left(
V\right) \gtrsim \Delta \left( V\right) $ makes the assumption of
an isolated qu2LS no longer valid. Panel (b) to the right of
Fig.~\ref{fig_3x3_tbeff} shows the effective pseudo-magnetic
fields defined via the most general effective Hamiltonian in the
qu2LS, namely $H\equiv a1+\mathcal{\vec{B}}\cdot \vec{\sigma}$,
where $\mathcal{\vec{B}}$ $\equiv \vec{B}_{{\rm eff}}$ stands for
the effective equivalent of a magnetic field. $\mathcal{B}_{z}$
can clearly be turned on and off by the gate voltages and ranges
from zero to the value limited by the applied voltages. Yet,
$\mathcal{B}_{x}$ is overwhelmingly set by the tunneling $t$ and
hardly responds to different applied voltages. In other words,
although we have access to a variety of gate voltages and diverse
ranges, the fixed $t$-value is the one that essentially determines
the $\sigma_x$ gate.  As $t$ cannot be varied, it negates the
qubit control one needs over the entire Bloch sphere.  The range
of $\mathcal{B}_x$ values obtained by varying local voltages is so
narrow that it is hardly visible in Fig.~\ref{fig_3x3_tbeff}b. In
order to see the difference between the maximum and minimum value
of $\mathcal{B}_{x}$ achieved, their difference $\left\vert \Delta
\mathcal{B}_{x}\right\vert $ is plotted separately: the first dip
in this curve is related to a change in sign in $\Delta
\mathcal{B}_{x}$, while the second kink is already well beyond the
two-level regime and originates in other higher lying states
taking over the ground state.

The numerical results show that despite having access to a large
set of voltage gates, the second qugate {\em cannot} be
implemented electrostatically under the assumptions of constant
tunneling and no real external magnetic field \cite{note}. One may
argue, that this happens because of the peculiar geometry chosen
and that there may be other geometries which would respond
differently. That this is not the case will be shown in the
following section.

\section{Electrostatic Interactions with Gates}

From an analytical point of view, some general statements can be
made on the charge states considered here. From our description of
the situation above, we can formulate the following two points:
\let\oldtheenumi\theenumi \let\oldlabelenumi\labelenumi
\renewcommand{\theenumi}{B\arabic{enumi}}
\renewcommand{\labelenumi}{\theenumi.}
\begin{enumerate}
 \item \label{stmt_qustate_separation} Encoding the qubit in the
charge and thus in the spatial wave function, demands that the
basis of the ground state pair be formed from two spatially
separated wave functions.
 \item \label{stmt_2level_compromise} Implementation of a
second qugate via electrostatic means with the tunneling $t$ kept
constant and with no real external magnetic field \cite{note}
results in compromising of the two-level low-energy system,
invalidating its use.
 \setcounter{oldenumi}{\value{enumi}}
\end{enumerate}

We can in fact demonstrate that these two points are true {\em in
general}, as evidenced by the specific geometry above. In order to
show this, we will use the following statements:
\renewcommand{\theenumi}{C\arabic{enumi}}
\renewcommand{\labelenumi}{\theenumi.}
\begin{enumerate}
\item \label{stmt_nodeless_gs} The ground state of any state
(single particle, singlet or triplet) for a real (not complex)
Hamiltonian must be nodeless, where for the states with more than
one particle one must consider the restricted space $\Omega \equiv
\vec{r}_{1}<\vec{r}_{2}<\ldots $ only within some unique sorting
scheme, where $\vec{r}_{i}$ points to the location of particle $i$
(this restriction is necessary since for example on the overall
space, the triplet states have an intrinsic node due to the
particle exchange symmetry).

\item \label{stmt_node_in_psi2}If a matrix element $\left\langle
\psi _{1}\right\vert V\left\vert \psi _{2}\right\rangle $ is $\neq
0$ for a local potential, then for $\psi _{1}\neq \psi _{2}$ at
least one of the two wave functions must have a node within the
space $\Omega $, and thus must be split off from the ground state
itself.
\end{enumerate}
\let\theenumi\oldtheenumi \let\labelenumi\oldlabelenumi

The argument for statement (\ref{stmt_nodeless_gs}) is similar to
one found in \cite{Slater53}. The statement follows from the
observation that any eigenstate $\psi \left(
\vec{r}_{1},\vec{r}_{2}\right) $ with a node within $\Omega $ has
a counterpart $\left\vert \psi \right\vert $ which has the same
energy expectation value $\int_{\Omega }\left\vert \psi
\right\vert \cdot H\cdot \left\vert \psi \right\vert
=E=\int_{\Omega }\psi H\psi $ and thus by the variational
principle, the ground state must be always nodeless or, at least,
can be chosen as such. Statement (\ref{stmt_node_in_psi2}) is
shown as follows: since $\int_{\Omega }\psi _{1}^{\ast }\left(%
\vec{r}_{1}, \vec{r}_{2}\right) V\left(%
\vec{r}_{1},\vec{r}_{2}\right) \psi _{2}\left(%
\vec{r}_{1},\vec{r}_{2}\right) \neq 0$ with the local potential
$V\left( \vec{r}_{1}^{~\prime },\vec{r}_{2}^{~\prime
};\vec{r}_{1},\vec{r}_{2}\right)
\equiv V\left( \vec{r}_{1},\vec{r}_{2}\right) \delta \left( \vec{r}%
_{1}^{~\prime }-\vec{r}_{1}\right) \delta \left( \vec{r}_{2}^{~\prime }-\vec{r}%
_{2}\right) $ there must some region in space where both $\psi
_{1}$ and $\psi _{2}$ are $\neq 0$ simultaneously. Yet, since
$\psi _{1}$ and $\psi _{2}$ are orthogonal eigenfunctions, in
order for $\langle \psi _{1}|\psi _{2}\rangle \ $to be $=0$ there
must be still another region in space with both $\psi _{1}$ and
$\psi _{2}$ unequal to zero but with a different sign compared to
the first region. Thus either $\psi _{1}$ or $\psi _{2}$ must
switch sign from one region to the other.

With this, statement (\ref{stmt_qustate_separation}) follows from
the observation that for some specific set of parameters (within
the working-range of the qubit) the ground state is degenerate;
thus utilizing statement (\ref{stmt_nodeless_gs}), both of these
ground states must be nodeless. Yet, they must be also orthogonal
to each other, and so similar to statement
(\ref{stmt_node_in_psi2}), $\psi _{1}$ and $\psi _{2}$ can be
chosen such that $\psi _{1}$ is $=0$ where $\psi _{2}\neq 0$ and
vice versa. This is what is meant by spatially separated wave
functions. Furthermore, since this ground state pair is supposed
to be sufficiently decoupled from the remaining states, this
situation may only change slightly during gate operations.

For statement (\ref{stmt_2level_compromise}), the Feshbach
formalism is employed once more. In the matrix representation, the
Hamiltonian of the qubit system with an isolated subspace with
index $\{1,2\}$ is given by
\begin{eqnarray}
H &=&\left(
\begin{array}{cc|ccc}
\varepsilon _{1}\left( \mathbf{V}_{g}\right) & 0 & \dots &
H_{1k^{\prime
}}\left( t\right) & \dots \\
0 & \varepsilon _{2}\left( \mathbf{V}_{g}\right) & \dots & \dots & \dots \\
\hline
\vdots & \vdots & \ddots & \vdots & H_{kk^{\prime }}\left( t\right) \\
H_{1k^{\prime }}^{\ast }\left( t\right) & \vdots & \dots &
\varepsilon
_{k}\left( \mathbf{V}_{g}\right) & \dots \\
\vdots & \vdots & H_{kk^{\prime }}^{\ast }\left( t\right) & \vdots
& \ddots
\end{array}%
\right)  \notag \\
&\equiv &%
\begin{pmatrix}
H_{PP}\left( \mathbf{V}_{g}\right) & H_{PQ}\left( t\right) \\
H_{PQ}^{+}\left( t\right) & H_{QQ}\left( t,\mathbf{V}_{g}\right)%
\end{pmatrix}
\label{H_PQ_rep}
\end{eqnarray}%
with $H_{PP}$ the projection of the Hamiltonian onto the $2D$
ground state space where $\varepsilon _{1}=\varepsilon _{2}$ for
$\mathbf{V}_{g}=0$ with $\mathbf{V}_{g}$ the set of external gate
potentials.  In this spatial representation, the potential
$\mathbf{V}_{g}$ enters only in the diagonal of the Hamiltonian
and, furthermore, the Hamiltonian is diagonal when $t=0$. So there
is no coupling of
$H_{PP}$ to the remaining space for $t=0$ since $H_{PQ}\left(%
0\right) =0$. From this structure of the Hamiltonian, the first
qugate ($\sigma _{z}$) is easily realized by choosing
$\mathbf{V}_{g}$ such that $\varepsilon _{1}\neq \varepsilon
_{2}$; the second qugate, however, must be realized through
coupling to the remaining space. For simplicity but without
restricting the case, a $\mathbf{V}_{g}$ is chosen that leaves
$\varepsilon _{1}$ and $\varepsilon _{2}$ constant or just shifts
them together uniformly; the effective two-level Hamiltonian
created by the Feshbach formalism effectively folds the remaining
Hilbert space into the reduced Hamiltonian $H_{PP}$ and thus
creates a shift in the diagonal, as well as generating the
off-diagonal elements. The latter terms can be straightforwardly
related to the $\sigma _{x}$ which gives the splitting in
Eq.~(\ref{delta_level_splitting}). Thus by comparison, the
effective second qugate for singlet states is approximated by
\begin{equation*}
\mathcal{B}_{x}\approx 32\frac{t^{4}}{\left[ \Delta \left( \mathbf{V}%
_{g}\right) \right] ^{3}}
\end{equation*}%
where $\Delta (\mathbf{V}_{g})\equiv \Delta _{0}+\Delta
\varepsilon (\vec{V }_{g})$ is the gap with applied
$\mathbf{V}_{g}$, and $\left\vert \Delta \varepsilon /\Delta
_{0}\right\vert \ll 1$, so that the second gate is approximated by
\begin{equation}
\mathcal{B}_{x}\approx 32\frac{t^{4}}{\Delta _{0}^{3}}\cdot \left(
1-\frac{3\Delta \varepsilon \left( \mathbf{V}_{g}\right) }{\Delta
_{0}}\right) \label{2nd_qugate}
\end{equation}%
Note that this is a maximum estimate since all relevant higher
lying energies are supposed to behave collectively. Yet, as seen
from Eq.~(\ref{2nd_qugate}), this second qugate has a much weaker
dependence on the gate voltages since it must be mediated by the
coupling $t$, while the first qugate is sensitive to the gate
voltages as $\Delta \varepsilon (\mathbf{V}_{g})$ directly. For
the ground state two-level system to be sufficiently ideal in the
sense of decoupled from the rest of the system, it must hold that
$t/\Delta _{0}\ll 1$ and also $\left\vert \Delta \varepsilon
/\Delta _{0}\right\vert \ll 1$ for the gate operation to not
interfere with the higher lying states. The consequence is that
the initial splitting for the singlet states is small and the
effect of the second qugate only changes this splitting by a
fraction which is an order of magnitude smaller. In order to get a
significant contribution, the second condition $\left\vert \Delta%
\varepsilon /\Delta _{0}\right\vert \ll 1$ would have to be lifted
but that obviously sacrifices the two-level system altogether.
This proves statement (\ref{stmt_2level_compromise}).

Local electrostatic interaction of voltage gates with a qubit
system is therefore not sufficient for a full set of single qubit
rotations. However, revision of the arguments brought forward
clearly leaves two ways out of this dilemma: first, the gap
($\mathcal{B}_{x}$) is controlled by the tunneling. Thus tunable
tunneling allows for the second qugate needed as is well-known
\cite{Gardelis03}. Second, the Hamiltonian was assumed to be real.
The argument of a nodeless ground state wave function very much
relies on that fact since for a real wave function a sign change
is only possible via a transition through zero while in the
complex case this is no longer required. With this, statement
(\ref{stmt_qustate_separation}) becomes irrelevant, and the
freedom on $\left\langle \psi _{1}\right\vert V\left\vert \psi
_{2}\right\rangle $ $\neq 0$ is greatly increased. Specifically,
an external magnetic field which makes the Hamiltonian complex,
will in fact not increase the initially existent (but constant!)
$\sigma _{x}$ qugate for singlet states, but it can reduce it to
zero. Eventually, this is again equivalent to an effectively
tunable tunneling \cite{note}.

\section{Conclusions} 

The use of spatial wave functions to encode quantum bits has been
analyzed with respect to capacitive electrostatic interactions.
With emphasis on systems such as lithographically grown arrays
where the tunneling is fixed to a great extent by the thickness of
the tunnel barriers (oxide layers) used, it was shown that with
constant tunneling $t$ and no external magnetic field present, the
single qubit operations from a system built this way are severely
limited. The analysis shows that a full set of single qubit
operations requires either a tunable tunneling or an external
magnetic field which makes the wave function complex and thus
introduces further flexibility. A uniform external magnet field
applied to one qubit, on the other hand, needs to be sufficiently
localized with respect to an ensemble of qubits eventually needed.
This is a very challenging experimental task. The parameters for
the dot-dot capacitances chosen give energy scales for the
tunneling operations in the constrained two-level system in the
few $\mu$eV range. Capacitances of about $45\unit{aF}$ are related
to dimensions of a few hundred $\unit{nm}$. Therefore, still
possible smaller sizes leave room for increased energy scales.

\begin{acknowledgements}
We acknowledge helpful discussions with J. Heremans and D.
Phillips, as well as support from NSF Grant NIRT 0103034, and the
Condensed Matter and Surface Sciences Program at Ohio University.
\end{acknowledgements}

\appendix

\section{Capacitance matrix and qudot interaction \label{appendix_C-Matrix}}

The total capacitance matrix of the system consisting of dots and
gates is written as

\begin{equation}
C_{tot}\equiv \left(
\begin{array}{cc}
C_{dot-dot} & C_{dot-gate} \\
C_{gate-dot} & C_{gate-gate}%
\end{array}%
\right) \equiv \left(
\begin{array}{cc}
C_{11} & C_{12} \\
C_{21} & C_{22}%
\end{array}%
\right)  \label{C-Matrix}
\end{equation}%
where the matrix has been decomposed into convenient block
notation which separates the dot-dot and dot-gate interactions.
For a network of capacitors connecting pairs of objects (dots or
gates), the individual matrix elements are given as

\begin{equation}
C_{ij}^{tot}=%
\begin{cases}
C_{i,0}+\sum_{j^{\prime }\neq i}^{dots,~gates}C(i\leftrightarrow
j^{\prime })
& \text{for $i=j$}, \\
-C(i\leftrightarrow j) & \text{$i\neq j$}.%
\end{cases}
\label{C-Matrix_ij}
\end{equation}%
where $C_{i,0}$ is the necessary self-capacitance of the $i$-th object and $%
C(i\leftrightarrow j)$ is the capacitor value between the two
objects $i$ and $j$. Here, the capacitance matrix is approximated
by a network of nearest neighbor capacitors where the specific
capacitance values enter as
parameters to the model. With the block matrix notation given in Eq.~(\ref%
{C-Matrix}), the coefficients $V_{ij}$ and $V_{i}$ for the Hubbard
Hamiltonian in Eq.~(\ref{Hubbard_H}) follow as

\begin{equation}
V_{ij}=e^{2}\left[ C_{11}^{-1}\right] _{ij},\ V_{i}=e~\,\left[ \mathbf{V}%
_{g}\cdot C_{21}C_{11}^{-1}\right] _{i}  \label{V_in_C-Matrix}
\end{equation}%
The interaction of the dots to the gates is mediated as expected
by the off-diagonal block $C_{21}$ of the total capacitance matrix
and is linear in
the applied set of gate voltages $\mathbf{V}_{g}=\{V_{g,1},V_{g,2},\ldots \}$%
.


\begin{thebibliography}{99}
\bibitem{note} It turns out it is also possible to control the
effective tunneling in the qudot array via the application of an
actual magnetic field.  We will discuss that possibility
elsewhere.
\end{thebibliography}
\end{document}